\documentclass[conference, table]{IEEEtran}
\IEEEoverridecommandlockouts

\usepackage{amsmath,amssymb,amsfonts,lipsum,makecell}
\usepackage{booktabs,multirow}
\usepackage{algorithmic}
\usepackage{graphicx}
\usepackage{textcomp}
\usepackage{xcolor}
\usepackage{stfloats}
\usepackage{hyperref}
\usepackage{arydshln}
\usepackage[square,sort&compress,comma,numbers]{natbib}

\usepackage[]{color-edits}
\addauthor{ma}{red}
\addauthor{ua}{blue}
\definecolor{gray_utku}{rgb}{0.4, 0.4, 0.4}
\newcommand{\var}[1]{\textcolor{gray_utku}{\text{\tiny$\pm$#1}}}
\def\x{{\mathbf x}}
\def\y{{\mathbf y}}
\def\E{{\mathbf E}_\Omega}
\def\n{{\mathbf n}}
\def\btheta{{\boldsymbol \theta}}

\begin{document}

\title{Edge Computing for Physics-Driven AI in Computational MRI: A Feasibility Study}
\author{\IEEEauthorblockN{Ya\c{s}ar Utku Al\c{c}alar}
\IEEEauthorblockA{\textit{College of Science and Engineering} \\
\textit{University of Minnesota}\\
Minneapolis, USA \\
ORCID 0009-0000-6413-2588}
\and
\IEEEauthorblockN{Yu Cao}
\IEEEauthorblockA{\textit{College of Science and Engineering} \\
\textit{University of Minnesota}\\
Minneapolis, USA \\
ORCID 0000-0001-6968-1180}
\and
\IEEEauthorblockN{Mehmet Ak\c{c}akaya}
\IEEEauthorblockA{\textit{College of Science and Engineering} \\
\textit{University of Minnesota}\\
Minneapolis, USA \\
ORCID 0000-0001-6400-7736}
}

\maketitle

\begin{abstract}
Physics-driven artificial intelligence (PD-AI) reconstruction methods have emerged as the state-of-the-art for accelerating MRI scans, enabling higher spatial and temporal resolutions. However, the high resolution of these scans generates massive data volumes, leading to challenges in transmission, storage, and real-time processing. This is particularly pronounced in functional MRI, where hundreds of volumetric acquisitions further exacerbate these demands. Edge computing with FPGAs presents a promising solution for enabling PD-AI reconstruction near the MRI sensors, reducing data transfer and storage bottlenecks. However, this requires optimization of PD-AI models for hardware efficiency through quantization and bypassing traditional FFT-based approaches, which can be a limitation due to their computational demands. In this work, we propose a novel PD-AI computational MRI approach optimized for FPGA-based edge computing devices, leveraging 8-bit complex data quantization and eliminating redundant FFT/IFFT operations. Our results show that this strategy improves computational efficiency while maintaining reconstruction quality comparable to conventional PD-AI methods, and outperforms standard clinical methods. Our approach presents an opportunity for high-resolution MRI reconstruction on resource-constrained devices, highlighting its potential for real-world deployment.
\end{abstract}


\section{Introduction}
Physics-driven artificial intelligence (PD-AI) reconstruction has emerged as the state-of-the-art method for accelerating magnetic resonance imaging (MRI) \cite{hammernik2018VarNet, schempler2018deep, aggarwal2019MoDL, knoll2020deep-survey, hosseini2020dense}. These methods are based on classical optimization algorithms for solving a regularized least squares objective function, where the regularizers, or their associated proximal operators, are learned via neural networks~\cite{hammernik2023SPM}. Importantly, data fidelity of the output with the acquired MRI data, which is collected in the frequency domain, is enforced via the known forward operator based on MRI physics using classical methods with learnable parameters~\cite{hammernik2023SPM}, which has been shown to improve performance and generalizability~\cite{knoll2020fastMRIchallenge-1,muckley2021fastMRIchallenge-2}.

\looseness=-1
In MRI, there is an inherent trade-off between spatiotemporal resolution, signal-to-noise-ratio (SNR) and imaging time \cite{akcakaya2022reconbook}. Thus, the faster imaging afforded by these PD-AI methods have enabled a trade-off for improved spatiotemporal resolutions in many applications~\cite{yaman2021_3D-LGE,demirel2023NER,gulle2023CAMSAP,gu2024_ISBI_MEfMRI}. In particular, for functional MRI (fMRI) of the human brain, a critical tool in neuroscientific studies and discoveries, PD-AI based reconstruction approaches have ushered in unprecedented sub-millimeter resolutions~\cite{demirel2023NER} and potential for higher temporal resolutions~\cite{demirel2021EMBC_20fold_7TfMRI,gulle2023CAMSAP}. In a typical fMRI experiment, hundreds of brain volumes are acquired over time at high temporal resolutions \cite{vizioli2021lowering}. With the advancements in spatial resolution afforded by PD-AI reconstructions, this translates to large amounts of raw data, especially since the data needs to be saved for multiple sensors, also known as acquisition coils \cite{pruessmann1999sense}, as complex-valued arrays, over many acquisition volumes. In turn, this creates problems for data transmission, data storage, computational loads and inference times. Thus, an alternative approach that may tackle these issues near the sensors prior to the console computer may be desirable for streamlining ultrahigh-resolution MRI at high acceleration rates.

Edge computing devices based on field-programmable gate arrays (FPGAs)~\cite{ma2018automatic, ma2018alamo, zhang2019caffeine, suh2023algorithm, nair2023fpga} offer a promising path for moving part or all of the PD-AI based computational MRI processing pipeline closer to the MRI sensors, thereby reducing data transmission and storage requirements substantially. These applications may range from data reduction techniques~\cite{buehrer2007array} to the actual PD-AI reconstruction algorithm~\cite{hammernik2023SPM}.

\looseness=-1
In this study, we sought to establish the feasibility of a PD-AI reconstruction algorithm suitable for FPGA implementation and validate its performance against its commonly used counterpart. In particular, we consider both quantization effects and elimination of fast Fourier transforms (FFTs) and inverse FFTs (IFFTs) between the image and frequency domains, the latter of which is the measurement domain in MRI. For the latter,  we focus on the case of equispaced sub-sampling with no calibration lines that is ubiquitous in echo-planar imaging based acquisitions such as fMRI~\cite{van2013wu_minn_HCP,akcakaya2022reconbook}. This allows to reformulate the data fidelity operation without the need for multiple FFTs and IFFTs, which is critical for FPGA implementations because the precision limitations can cause significant accuracy loss. For the former, we assess how 8-bit quantization in the convolutional neural network (CNN), suitable for FPGA implementation, affects the performance of the PD-AI neural network. Our results show that PD-AI methods can be implemented with sufficient quantization and without repeated FFTs/IFFTs, suitable for using FPGAs on edge computing devices such as the sensors, for many high-resolution MRI applications that lead to a large amount of raw data.

\section{Methods}

\subsection{MRI Acquisition Model} \label{sec:2a}
MRI acquisitions collect data in the frequency domain, also referred to as the k-space. This is done using multiple receiver coils, each of which is sensitive to different parts of the imaging volume \cite{pruessmann1999sense}. Due to the aforementioned resolution, scan time and SNR trade-off, MRI acquisitions are typically performed by sub-sampling the data, i.e. taking fewer measurements at a sub-Nyquist rate \cite{akcakaya2022reconbook}. In this case, the forward model is given as:
\begin{equation} \label{Eq:Forward_Model_MultiCoil_Percoil}
    \y_\Omega^k = {\bf P}_{\Omega}\mathcal{F}_N{\bf C}^{k}\x + \n^k, \:\: k \in \{1, \dots, n_c\}, 
\end{equation}
where ${\bf x} \in {\mathbb C}^N$ is the (vectorized) image of interest, ${\bf y}_\Omega^k \in {\mathbb C}^M$ is the (vectorized) sub-sampled MRI data acquired at the $k^\textrm{th}$ coil, ${\bf P}_{\Omega}$ is an $M \times N$ masking operator that samples the k-space points specified by $\Omega$, $\mathcal{F}_N$ is the $N$-point FFT, ${\bf C}^{k} \in {\mathbb C}^{N \times N}$ is a diagonal matrix specifying the local sensitivity of the $k^\textrm{th}$ coil, ${\bf n}^k \in {\mathbb C}^M$ is the thermal measurement noise for the $k^\textrm{th}$ coil, and $n_c$ is the number of coils \cite{pruessmann1999sense}. This system is often abbreviated by concatenating across the coil dimension for a more compact representation:
\begin{equation} \label{Eq:Forward_Model_MultiCoil}
    \y_{\Omega} = \E\x + \n, 
\end{equation}
 where $\y_{\Omega}$ is the acquired sub-sampled k-space data across all coils, $\E$  
 is the forward encoding operator that concatenates all the coil-based encoding operators in \eqref{Eq:Forward_Model_MultiCoil_Percoil}, i.e. ${\bf P}_{\Omega}\mathcal{F}_N{\bf C}^{k}$, across $k \in \{1, \dots, n_c\}.$

\subsection{Physics-Driven AI (PD-AI) in Computational MRI} \label{sec:2b}
The associated inverse problem for computational MRI solves a regularized least squares objective function:
\begin{equation} \label{Eq:rls_obj}
\underset{\mathbf{x}}{\text{argmin}} \left\| \mathbf{y}_\Omega - \mathbf{E}_\Omega \mathbf{x} \right\|_2^2 + \mathcal{R}(\mathbf{x}).
\end{equation}
In this objective function, the first term imposes data fidelity of the estimate with respect to the acquired k-space data. The second term, $\mathcal{R}(\cdot)$, is a regularizer for the typically ill-conditioned first term. There are numerous iterative optimization algorithms for solving \eqref{Eq:rls_obj}, such as proximal gradient descent or variable splitting \cite{fessler2020SPM}. These methods typically alternate between a data fidelity operation to ensure consistency with the acquired data and a proximal operator for $\mathcal{R}(\cdot)$ \cite{fessler2020SPM}.

In PD-AI methods, a common approach is to unroll/unfold such an iterative optimization algorithm for a fixed number of steps \cite{hammernik2023SPM}. This unrolled network is trained end-to-end, where both the proximal operator for the regularizer and the weights for the linear data fidelity operation, are learned jointly. This can be done in a supervised setup, where the network output is compared to a ground-truth reference data \cite{hammernik2018VarNet,schempler2018deep}, or in an unsupervised/self-supervised manner~\cite{yaman2020SSDU,yaman2022mmssdu,yaman2022zeroshot,akcakaya2022_SPMsurvey,zhang2024ccssdu,alcalar2024_ISBI,alcalar2025SPIC_SSDU}. %\macomment{also cite your ISBI paper + ICCV arxiv}.

In this work, we consider the unrolling of the variable splitting with quadratic penalty (VSQP) algorithm, implemented in other prior studies \cite{hosseini2020dense, gu2022revisiting, demirel2023SIIM}. Succinctly, VSQP solves two sub-problems to optimize \eqref{Eq:rls_obj} as follows:
\begin{subequations}
    \begin{equation}
        \mathbf{z}^{(i)} = \arg\min_{\bf{z}} \|\mathbf{x}^{(i-1)} -\mathbf{z}\|_2^2 + \mathcal{R}(\mathbf{z}), \label{eq:vsqp_1}
    \end{equation}
    \begin{equation}
        \mathbf{x}^{(i)} = \arg\min_{\bf x} \|\mathbf{y}_\Omega - \mathbf{E}_\Omega \mathbf{x}\|_2^2 + \mu \|\mathbf{x} - \mathbf{z}^{(i)}\|_2^2, \label{eq:vsqp_2}
    \end{equation}
\end{subequations}
where (\ref{eq:vsqp_1}) corresponds to the proximal operator for ${\mathcal R}(\cdot)$ and is solved implicitly by a neural network. On the other hand, (\ref{eq:vsqp_2}) is the data fidelity operation, which is a least squares term with a well-known closed-form solution:
\begin{equation}
    \mathbf{x}^{(i)} = \left( \mathbf{E}^H_{\Omega} \mathbf{E}_{\Omega} + \mu \mathbf{I} \right)^{-1} (\mathbf{E}^H_{\Omega} \mathbf{y}_\Omega + \mu \mathbf{z}^{(i)}).\label{eq:closed_form}
\end{equation}

In our baseline PD-AI implementations, the VSQP algorithm is unrolled for 10 unrolls. The regularizer is implemented using a CNN based on a ResNet architecture~\citep{timofte2017ntire} with 15 residual blocks, which was successfully used in prior studies \cite{hosseini2020dense, gu2022revisiting, demirel2021EMBC_20fold_7TfMRI}. 
To avoid the matrix inversion in (\ref{eq:vsqp_2}) for data fidelity, this sub-problem itself was solved using an unrolled conjugate gradient (CG) method with 10 iterations~\cite{aggarwal2019MoDL}. The network was trained end-to-end over 100 epochs using the normalized $\ell_1$-$\ell_2$ loss in a supervised manner:
\begin{equation}
    \mathcal{L}(\x_\text{ref}, \hat{\x}) = \frac{\|\x_\text{ref} - \hat{\x}\|_2}{\|\x_\text{ref}\|_2} + \frac{\|\x_\text{ref} - \hat{\x}\|_1}{\|\x_\text{ref}\|_1},
    \label{eq:loss_function}
\end{equation}
where $\hat{\x}$ is the network output, $f(\x_\Omega,\E;\btheta)$. 

\begin{figure}[t]
    \centerline{\includegraphics[width=\columnwidth]{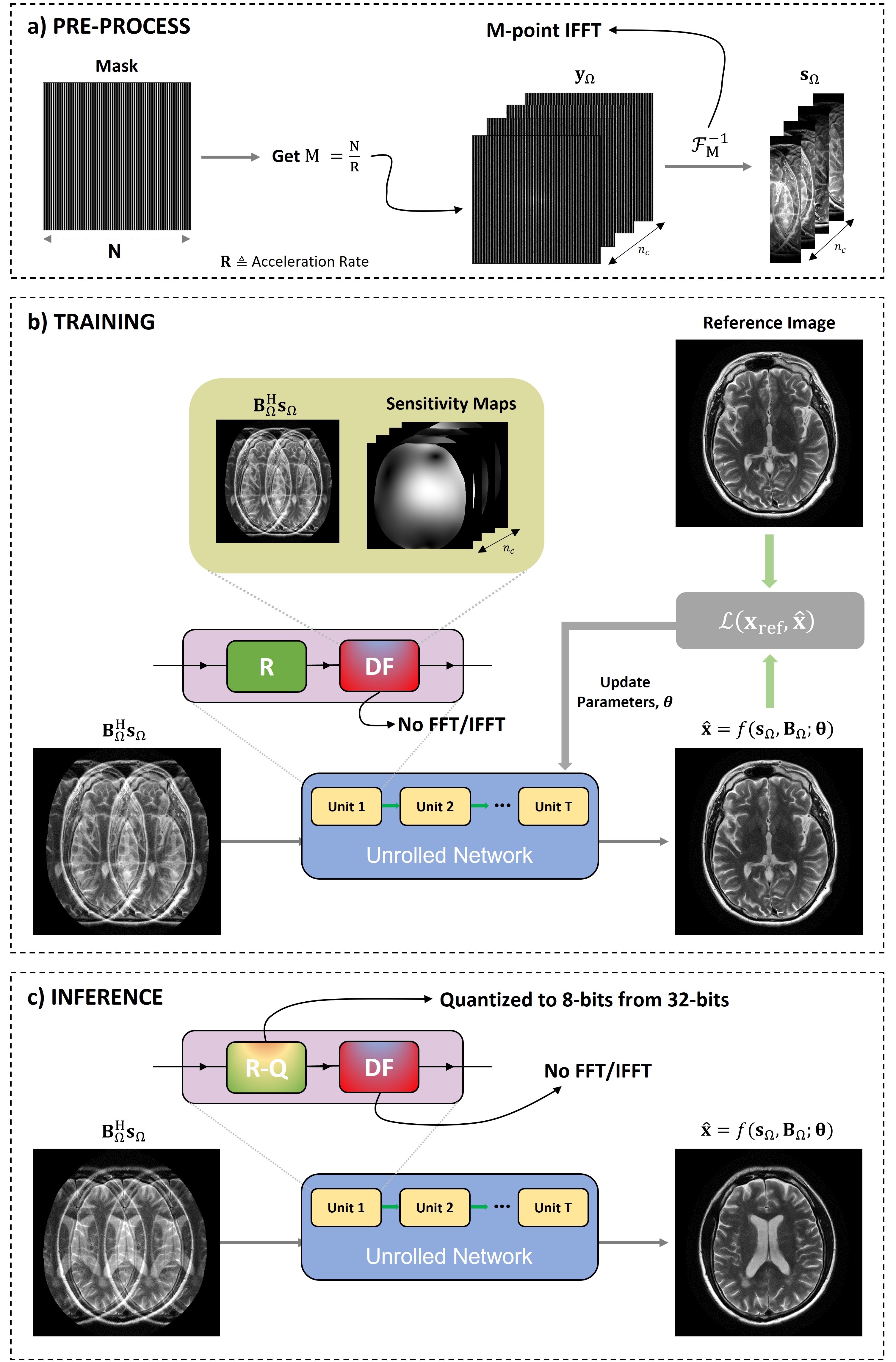}}
    \caption{Proposed edge computing amenable PD-AI. (a) Initially, the acquired k-space data undergoes pre-processing based on the acquisition mask. This is the sole instance in the entire pipeline where an FFT/IFFT operation is performed, and it is executed only once per slice. (b) Subsequently, an unrolled network, consisting of a neural network that implements the proximal operator of a regularizer (R) and a data fidelity (DF) unit that does not use any FFT/IFFTs, is trained in a supervised manner using 32-bit data. (c) During inference, 8-bit quantization is applied to the regularizer (R-Q), significantly reducing memory requirements while maintaining reconstruction accuracy.}
    \label{fig:methods}
\end{figure}

\begin{figure*}[t]
    \centerline{\includegraphics[width=\textwidth]{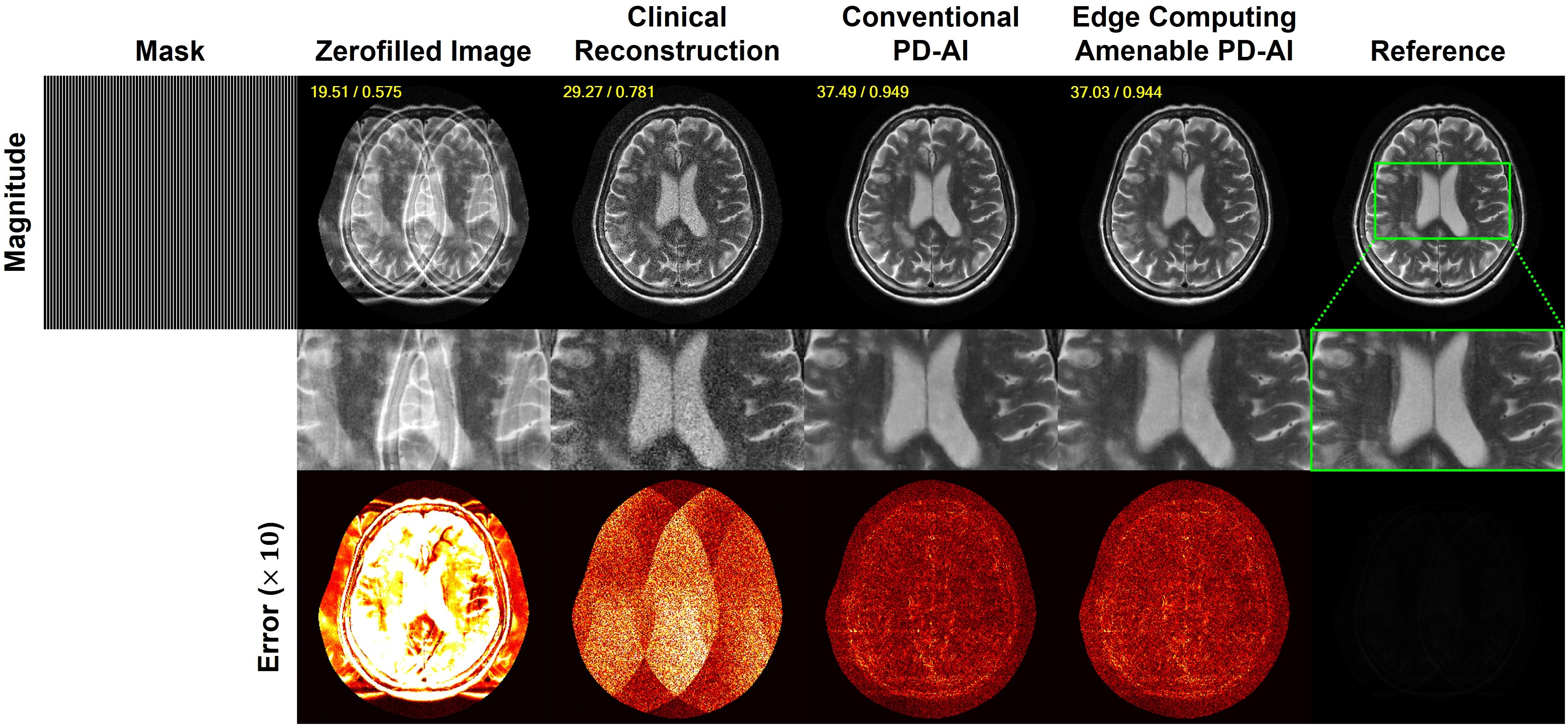}}
    \caption{Comparison of clinical reconstruction, conventional PD-AI, and the proposed edge computing-amenable PD-AI strategies. For a representative slice, the results demonstrate that the proposed method significantly enhances clinical reconstruction, maintaining comparable quality to conventional PD-AI, despite utilizing 8-bit complex data rather than the standard 32-bit complex data for the CNN. The clinical reconstruction shows noise amplification at high acceleration rates, which is mitigated by the the PD-AI approaches.}
    \label{fig:results}
\end{figure*}

\subsection{Proposed PD-AI Amenable to Edge Computing}

In order to ensure that the PD-AI reconstruction algorithm can be implemented in edge computing devices, for instance using FPGAs on the MRI coil arrays, care must be taken in two aspects. First, the effect of quantization on the network from 32-bit complex-valued processing to 8-bit complex-valued processing needs to be investigated. Second, repeated applications of FFTs and IFFTs in \eqref{eq:closed_form} for calculating $\mathbf{E}^H_{\Omega} \mathbf{E}_{\Omega}$ needs to be avoided. As stated, the AI-PD implementation in Section \ref{sec:2b} requires the calculation of $\mathbf{E}^H_{\Omega} \mathbf{E}_{\Omega}$ in 10 iterations of the CG in each of the 10 unrolls. This corresponds to $100 \cdot n_c$ FFTs and $100 \cdot n_c$ IFFTs, which is prohibitive for FPGA implementations, in addition to the $n_c$ IFFTs required to calculate $\mathbf{E}^H_{\Omega} {\bf y}_\Omega$.

Instead, we note that the most commonly used sub-sampling pattern $\Omega$ for high-volume high-resolution acquisitions considered here, such as fMRI, is an equispaced sub-sampling pattern with no calibration region \cite{van2013wu_minn_HCP}. Here, equispaced sub-sampling for acceleration rate $R = N/M$ means that every $R^{\textrm{th}}$ k-space line is sampled in the phase encode dimension \cite{pruessmann1999sense}. In this case, the k-space data acquired in the $k^\textrm{th}$ coil can be transformed to image domain for processing \cite{pruessmann1999sense}, as follows:
\begin{equation} \label{imdomain_meas}
    {\bf s}^k = %\frac{1}{\sqrt{R}} \cdot 
    {\cal F}_M^{-1} {{\bf y}_\Omega^k}, \:\: k \in \{1, \dots, n_c\}, 
\end{equation}
where ${\cal F}_M^{-1}$ is an $M$-point IFFT. For acceleration rate $R = N/M$, it is easy to see that this corresponds to an $R$-fold foldover of the original image of interest in that coil, ${\bf C}^k {\bf x}$, as an aliasing artifact \cite{pruessmann1999sense}, as seen from Fig.~\ref{fig:methods}a. In other words, the corresponding forward operator for the coil can be rewritten in image domain as ${\bf R}_\Omega = \sqrt{R}{\cal F}_M^{-1} {\bf P}_{\Omega} {\cal F}_N$, which is an $M \times N$ matrix, whose $k^{th}$ row has zeros in all coordinates except for coordinates specified by $\{k, k+R, \dots, k +(R-1)R\}$, which are equal to 1. We emphasize that ${\bf R}_\Omega$ can be calculated directly in image domain without any FFT or IFFT. Furthermore, note that with appropriate scaling in ${\bf R}_{\Omega}$, it can be exactly ensured that $||\y_{\Omega} - \E\x||_2^2 = ||{\bf s}_{\Omega} - {\bf B}_{\Omega}{\bf x}||_2^2$.

Finally, as in Section \ref{sec:2a}, the system in image domain can be abbreviated by concatenating across the coil dimension as:
\begin{equation}
    {\bf s}_{\Omega} = {\bf B}_{\Omega}{\bf x} + {\bf n},
\end{equation}
where $s_{\Omega}$ is the acquired sub-sampled k-space data across all coils, ${\bf B}_{\Omega}$ is the image-domain forward encoding operator, which concatenates all coil-based ${\bf R}_\Omega{\bf C}^k$, across $k \in \{1, \dots, n_c\}$, \emph{without requiring} any FFT or IFFT calculation.

This edge computing amenable PD-AI algorithm was implemented as follows: 
For the 8-bit quantization of the CNN, both the activations and weights were quantized to 8 bits using a per-tensor affine quantization scheme~\cite{xiao2023smoothquant}, rather than applying quantization only to the weights. In this approach, each tensor is scaled and shifted independently, allowing for a more flexible representation of the data. While more challenging in terms of performance, this method, in contrast to weight-only quantization, enables improved memory efficiency and computational performance, particularly for deployment on resource-constrained devices. By reducing precision from 32-bit to 8-bit, this approach achieves an approximate \emph{$4\times$ reduction in memory usage}, making it significantly more feasible for edge computing.

Notably, we did not perform quantization-aware training (QAT)~\cite{jacob2018quantization} but instead applied quantization to a pre-trained network that originally used 32-bit complex data. This approach is reflective of a more realistic scenario, as many existing MRI reconstruction models are not typically trained with quantization in mind, but rather with high-precision data. Furthermore, as running inference with quantized networks on GPUs is still an area of ongoing development, we focused on performing inference on the CPU for this study.

\vspace{4mm}
\section{Experimental Evaluations}
\subsection{Imaging Experiments}
The proposed edge computing amenable PD-AI algorithm was comprehensively evaluated against the conventional PD-AI algorithm both qualitatively and quantitatively using raw MRI k-space data. Quantitative evaluation was performed using peak signal-to-noise ratio (PSNR) and structural similarity index (SSIM). Imaging experiments were conducted on the fully-sampled multi-coil brain MRI data from the New York University (NYU) fastMRI database~\cite{zbontar2019fastmri_dataset-arXiv,knoll2020fastmri_dataset-journal}. In particular, axial T2-weighted (ax T2-W) images with a matrix size of 320$\times$320 were used. These datasets were collected with $n_c$=16 receiver coils. Datasets were retrospectively undersampled using an equidistant pattern with acceleration rates of $R=4$ without a central calibration region. Note that we used axial T2-weighted images instead of fMRI for two reasons. First, fMRI data are almost always collected with sub-sampling~\cite{van2013wu_minn_HCP} thus there is no ground-truth reference, against which one can perform quantitative comparisons \cite{demirel2021EMBC_20fold_7TfMRI}. Second, axial T2-weighted images are higher resolution than typical fMRI acquisitions, aligning the evaluations more closely with the goal of enabling higher resolution brain scans, while also testing the method for a larger matrix size.

\subsection{Results}

\begin{table}[t]
    \setlength{\tabcolsep}{4.5pt}
    \caption{Quantitative comparison of clinical reconstruction, conventional PD-AI, and the proposed edge computing amendable PD-AI across different metrics.}
    \begin{center}
    \begin{tabular}{@{}p{3.5cm}ccccc@{}}
    \hline
    \addlinespace[3pt]
    \makecell[c]{Method} & PSNR$\uparrow$ & SSIM$\uparrow$ & \makecell[c]{Inference Time \\ on CPU (s)}\\ %& \makecell[c]{Network \\ Size (GB)}\\
    \addlinespace[2pt]
    \hline
    \addlinespace[2pt]
    \makecell[c]{Clinical Recon-\\ struction~\cite{pruessmann1999sense,pruessmann2001cgsense}} & 29.05\var{2.68} & 0.816\var{0.062} & 0.25\\
    \addlinespace[2pt]
    \makecell[c]{Conventional \\ PD-AI~\cite{hammernik2018VarNet,aggarwal2019MoDL}} & 35.68\var{2.65} & 0.932\var{0.042} & 4.92\\
    \addlinespace[2pt]
    \arrayrulecolor{gray} \hdashline
    \addlinespace[2pt]    
    \makecell[c]{Edge Computing \\ Amendable PD-AI {\bf (Ours)}} & 35.21\var{2.44} & 0.924\var{0.037} & 2.49\\
    \addlinespace[2pt]
    \arrayrulecolor{black} \hline
    \end{tabular}
    \vspace{-5mm}
    \label{tab:table}
    \end{center}
\end{table}

We compared the performance and speed of our method with conventional PD-AI~\cite{hammernik2018VarNet, aggarwal2019MoDL, hammernik2023SPM} and parallel imaging reconstruction~\cite{pruessmann1999sense, pruessmann2001cgsense}, the latter of which is widely used in clinical practice. Fig.~\ref{fig:results} compares the conventional and the proposed edge computing amenable PD-AI strategies, as well as clinical reconstruction. As expected, clinical reconstruction exhibits noise amplification at high acceleration rates. The proposed approach significantly improves upon clinical reconstruction while maintaining comparable quality to conventional PD-AI, despite using 8-bit complex data instead of 32-bit complex data for the CNN.

Table~\ref{tab:table} presents a quantitative comparison between the clinical reconstruction, conventional PD-AI, and the proposed edge computing amenable PD-AI, reinforcing the observations from the visual results. The proposed edge computing amenable PD-AI shows similar performance to the conventional PD-AI in terms of reconstruction quality, with only a slight decrease in PSNR and SSIM values. However, it offers a notable improvement in inference time, running significantly faster on the CPU (AMD EPYC 7352 24-Core Processor). The significant reduction in inference time is largely due to the use of 8-bit complex data, which reduces the computational burden. This suggests that the proposed approach, optimized for 8-bit data, is well-suited for deployment on resource-constrained edge devices like FPGAs, where both performance and real-time processing are important. Our proposed approach provides a favorable trade-off between computational efficiency and reconstruction quality, making it a suitable candidate for FPGA-based applications.

\section{Discussion and Conclusion}
In this study, we introduced a novel approach for implementing PD-AI reconstruction in FPGA-based edge computing devices. Our feasibility study shows a potential solution in addressing the challenges associated with the ever increasing data sizes in high-resolution MRI. Our technical developments focused on the elimination of multiple FFTs/IFFTs and the impact of quantization on the network performance. Our results demonstrated that the proposed method offers comparable reconstruction quality to conventional PD-AI approaches while significantly improving computational efficiency, making it suitable for edge computing applications.

Further developments may involve exploring 4-bit quantization and the quantization of the data fidelity unit. While quantization of the data fidelity unit presents challenges due to the limited arithmetic operations in current deep learning frameworks, advancements in these fields may make it more tractable, potentially paving the way for coil quantization in MRI as well.  However, these aspects were not the primary focus of the current study and will be explored in future research.

\section*{Acknowledgment}
This work was partially supported by NIH R01EB032830 and NIH P41EB027061.

\fontsize{8}{10}\selectfont
\bibliographystyle{IEEEbib}
\bibliography{refs}

\end{document}